\documentclass[10pt,a4paper]{article}

\usepackage{amssymb}
\usepackage{amsmath}
\usepackage{bm}
\usepackage{graphicx}
\usepackage{subfigure}
\usepackage{hyperref}
\usepackage{color,soul}

\usepackage{booktabs}
\usepackage{multirow}
\usepackage{array}
\usepackage{enumitem}
\usepackage{cite}
\usepackage[affil-it,auth-lg]{authblk}
\usepackage{eso-pic}

\usepackage{enumitem}

\begin{document}
	
	\title{Anisotropic Neutron Stars Modelling: Constraints in Krori-Barua Spacetime}
	\author[1]{Zacharias Roupas}
	\author[1]{Gamal G. L. Nashed}
	\affil[1]{Centre for Theoretical Physics, The British University in Egypt, Sherouk City 11837, Cairo, Egypt} 
	
	\date{\vspace{-5ex}}
	
	\maketitle

\begin{abstract}
	Dense nuclear matter is expected to be anisotropic due to effects such as solidification, superfluidity, strong magnetic fields, hyperons, pion-condesation. Therefore an anisotropic neutron star core seems more realistic than an ideally isotropic one. We model anisotropic neutron stars working in the Krori-Barua (KB) ansatz without preassuming an equation of state. We show that the physics of general KB solutions is encapsulated in the compactness. Imposing physical and stability requirements yields a maximum allowed compactness $2GM/Rc^2 < 0.71$ for a KB-spacetime. 
	We further input observational data from numerous pulsars and calculate the boundary density.  
We focus especially on data from the LIGO/Virgo collaboration as well as recent independent measurements of mass and radius of miilisecond pulsars with white dwarf companions by the \textit{Neutron  Star  Interior  Composition  Explorer} (NICER).
	For these data the KB-spacetime gives the same boundary density which surprisingly equals the nuclear saturation density within the data precision. Since this value designates the boundary of a neutron core, the KB-spacetime applies naturally to neutron stars. For this boundary condition we calculate a maximum mass of 4.1 solar masses.  
\end{abstract}

%%%%%%%%%%%%%%%%%%%%%%%%%%%%%%%%%%%%%%%%%%%%%%%%%%%%%%%%%%%%%%%%%
\section{Introduction}

Very early in the study of pulsars it was realized that anisotropies inside the star can grow due to superfluidity \cite{1969Natur.224..673B,1970PhRvL..24..775H} (see \cite{2019EPJA...55..167S} for a modern review) and solidification \cite{1971NPhS..231..145A,1972NPhS..236...37C,1973PhRvL..30..999C,1973NPhS..243..130S}. An anisotropic core may also originate \cite{2007ASSL..326.....H} from hyperons \cite{1998PhRvC..57..409B}, quarks \cite{1984PhR...107..325B} as well as pion and kaon condensates \cite{1972PhRvL..29..382S,1995PThPh..94..457T}. In addition, nuclear matter in a magnetic field becomes anisotropic, with different pressures in directions along and transverse to the field \cite{2010PhRvC..82f5802F,2019Univ....5..104F}. The electromagnetic energy-momentum tensor is naturally anisotropic. Accounting for all these theoretical predictions it seems more realistic that pulsars contain an anisotropic core, rather than an ideally isotropic one.

The theory of anisotropic compact objects in General Relativity has been developing for half a century. Bowers \& Liang \cite{Bowers:1974tgi} calculated the anisotropic generalization of Tolman-Oppenheimer-Volkov equation and generalized Bondi's analysis \cite{1964RSPSA.282..303B}. An isotropic solution involves the emergence of a tangential pressure $p_t = p_\theta = p_\phi$ in the angular directions that is different than the radial pressure $p_r \neq p_t$. If the anisotropy parameter is positive $\Delta \equiv p_t-p_r>0$ an additional repulsive anisotropic force enhances stability, enabling more compact stable configurations to appear in the anisotropic than in the isotropic case \cite{2002PhRvD..65j4011I}. It is in particular proposed \cite{GLEISER_2004,Bohmer2006} that anisotropic compact stars may be arbitrarily compact up to compactness $C=2GM/Rc^2$ equal to one.
Heintzmann \& Hillebrandt \cite{1975A&A....38...51H} estimated the maximum mass of anisotropic compact stars $M_{\rm max}\sim4M_\odot$ by use of semi-realistic equations of state. 
The Jeans stability criterion has been extended in the anisotropic case by Herrera \& Santos \cite{1995ApJ...438..308H}.

A lot of anisotropic solutions and anisotropic compact star models have been proposed and studied \cite{PhysRevD.26.1262,PhysRevD.77.027502,Thirukkanesh_2008,2016Ap&SS.361..339S,Maurya_2016,Maurya_2017,2018EPJC...78..673E,2019EPJC...79..885T,2019EPJC...79..853D,2019EPJP..134..600E,2019EPJC...79..138B}. Ivanov has calculated general bounds on the redshift for any anisotropic compact star in Ref. \cite{Ivanov_2002}.
We will work here in a metric ansatz introduced by Krori \& Barua \cite{1975JPhA....8..508K}. Anisotropic compact star models in the Krori-Barua spacetime in General Relativity have been studied in Refs. \cite{PhysRevD.82.044052,2012EPJC...72.2071R,Kalam_2013,Bhar:2014mta,2015Ap&SS.356..309B,Bhar_etaL_2015}
 and in modified theories of gravity in Refs. \cite{2015Ap&SS.359...57A,2016Ap&SS.361....8Z,2016Ap&SS.361..342Z,2016CaJPh..94.1024S,2017Ap&SS.362..237I,2018EPJC...78..307Y,2018IJGMM..1550093S,2019CoTPh..71..599F,2019EPJC...79..919S,2020IJMPA..3550013S,2020MPLA...3550354S}. 
 
 We will perform a general analysis of physical viability and stability of anisotropic solutions in KB-spacetime without preassuming an equation of state. We introduce dimensionless variables in which the KB-solutions can be parametrized with respect to the compactness. We impose general conditions for stability and physical consistency, which imply constraints on the maximum allowed compactness. We further use pulsars' observational data of total mass and radius to estimate the boundary density of the core within our model and calculate the predicted mass-radius curve under a certain boundary condition.

To this end we have to use measurements of the mass and radius of pulsars, which are independent and do not rely on pre-assumptions regarding the equation of state in the core. 
We shall use six pulsars, members of low-mass X-ray binaries, which present thermonuclear bursts and therefore is possbile to get correlated $M-R$ constraints \cite{2006Natur.441.1115O,2016ARA&A..54..401O}. More importantly we shall use data from additional two rotation-powered millisecond pulsars, PSR J0437-4715 and PSR J0030+0451, for which there exist reliable measurements of their radius independent from measurements of their mass and also not depending on assumptions regarding the equation of state
\cite{2016ARA&A..54..401O,2019MNRAS.490.5848G,2019ApJ...887L..25B,2019ApJ...887L..24M}. These two pulsars are special in that such measurements are as yet very rare. Both pulsars data are found to be consistent with the same boundary density which amazingly turns out to equal the nuclear saturation density that typically designates the boundary of the neutron core. 

In addition, we shall use data regarding the recent gravitational-wave signals GW170817 \cite{TheLIGOScientific:2017qsa,PhysRevLett.121.161101} and GW190814 \cite{2020ApJ...896L..44A}. We will finally supplement our analysis with data regarding quiescent low-mass X-ray binaries \cite{2006ApJ...644.1090H,2007ApJ...671..727W} and pulsars presenting thermonuclear bursts \cite{2016ApJ...820...28O}.

In the next section we review KB-spacetime and introduce our dimensionless variables. In section \ref{sec:phys_an} we perform the physics and stability analysis. In section \ref{sec:obs} we discuss observational data and in the final section we discuss our conclusions.

%%%%%%%%%%%%%%%%%%%%%%%%%%%%%%%%%%%%%%%%%%%%%%%%%%%%%%%%%%%%%%%%%
\section{Krori-Barua Spacetime}\label{sec:model}

A general spherically symmetric metric in General Relativity may be written in the spherical coordinates $(t,r,\theta, \phi)$ as
\begin{equation}
ds^2=-e^{\alpha(r)} \, c^2 dt^2+e^{\beta(r)}dr^2+r^2(d\theta^2+\sin^2\theta d\phi^2)\,,\label{eq:met1}
\end{equation}
where $\alpha(r)$ and $\beta(r)$ are unknown functions of $r$. The spherically symmetric, anisotropic energy momentum tensor may be written as
\begin{equation}\label{eq:T_enmom}
T_\nu^\mu{} =(\frac{p_t}{c^2}+\rho )u^\mu u_\nu + p_t\delta_a{}^\mu + (p_r - p_t) \xi^\mu \xi_\nu ,
\end{equation}
where $\rho$, $p_r$, $p_t$ denote the mass density, the radial pressure and the tangential pressure, respectively. We denote $u^\mu$ the four-velocity and $\xi^\mu$ is the unit space-like vector in the radial direction. The energy-momentum tensor can always be brought in the diagonal form $T^\mu_\nu = {\rm diag}(\rho c^2,-p_r,-p_t,-p_t)$.

Einstein equations give
\begin{align} 
\label{eq:f_rho}
 \frac{8\pi G}{c^2}\rho &= \frac{e^{-\beta}}{r^2} 
 \left( e^{\beta}+\beta'r-1 \right)\,,
 \\
\label{eq:f_p_r}
\frac{8\pi G}{c^4} p_r &= \frac{e^{-\beta}}{r^2}
\left( 1-e^{\beta}+r\alpha'\right)\,,
\\
\label{eq:f_p_t}
\frac{8\pi G}{c^4} p_t &= e^{-\beta}
\left( \frac{\alpha''}{2} - \frac{\alpha'\beta'}{4} + \frac{\alpha'^2}{4} + \frac{\alpha' - \beta'}{2r} \right)\,,
\end{align}
where prime denotes derivative w.r.t the radial coordinate $r$. 

Following Krori \& Barua \cite{1975JPhA....8..508K} we assume the following ansatz for the metric potentials 
\begin{align}\label{eq:pot}
\alpha(x) =a_0 x^2+a_1\,,\quad 
\beta(x) =a_2 x^2\,,
\end{align}
where however we use the dimensionless variable
\begin{equation}\label{eq:x_dless}
x\equiv \frac{r}{R} \in [0,1]
\end{equation}
and the star is assumed to be extended up to the radius $r=R$.
The parameters $a_0$, $a_1$, $a_2$ in our ansatz are dimensionless and will be determined from the matching conditions on the boundary. The KB ansatz (\ref{eq:pot}) ensures that the gravitational potentials and their derivatives are finite at the center.

We further introduce the characteristic density
\begin{equation}\label{eq:rho_star}
\rho_\star \equiv \frac{c^2}{8\pi G R^2}
\end{equation}
which we use to scale the density and pressures, getting the dimensionless variables
\begin{equation}\label{eq:rho+p_dless}
\tilde{\rho} = \frac{\rho}{\rho_\star}\,,\;
\tilde{p}_r = \frac{p_r}{\rho_\star c^2}\,,\;
\tilde{p}_t = \frac{p_t}{\rho_\star c^2}\,,\;
\tilde{\Delta} = \frac{\Delta}{\rho_\star c^2}\,.\;
\end{equation}
Here $\Delta(r) = p_t - p_r$ is the anisotropic parameter of the star.
Note that for a typical radius of neutron stars $R=10{\rm km}$ we get $\rho_\star = 5.4\cdot 10^{14}{\rm gr/cm^3}$ that is twice the nuclear saturation density $\rho_{\rm sat} = 2.7\cdot 10^{14}{\rm gr/cm^3}$.
Using the dimensionless variables (\ref{eq:rho+p_dless}) and substituting (\ref{eq:pot}) in the system (\ref{eq:f_rho})-(\ref{eq:f_p_t}) we get 
\begin{align}
\label{eq:rho_a}
\tilde{\rho} &= 
\frac{ e^{-a_2 x^2}}{x^2} \left( e^{a_2 x^2}-1 + 2a_2 x^2 \right)
\,, 
\\
\label{eq:p_r_a}
\tilde{p}_r &= 
\frac{ e^{-a_2 x^2}}{x^2} \left( 1-e^{a_2 x^2}+2a_0 x^2 \right)\,,
\\
\label{eq:p_t_a}
\tilde{p}_t &= 
e^{-a_2x^2}
\left( 2a_0  -a_2 + a_0 (a_0 - a_2) x^2 \right)\, ,
\\
\label{eq:Delta_a}
\tilde{\Delta} &= 
\frac{e^{-a_2 x^2}}{x^2} 
\left(  e^{a_2 x^2} - 1 - a_2 x^2 + a_0(a_0-a_2) x^4 \right) \,.
\end{align}

The mass contained within a radius $r$ of the sphere
is defined as
\begin{equation}
\label{eq:mas_def}
\mathcal{M}(r)={\int_0}^r 4\pi \rho \xi^2 d \xi \,.
\end{equation}
Substituting the density we get
\begin{equation}
\label{eq:mas_a}
\mathcal{M}(x)=M\, C^{-1} x\left( 1-e^{-a_2 x^2}\right) \, .
\end{equation}
where $M$ denotes the total mass of the star and $C$ is the compactness
\begin{equation}
C = \frac{2GM}{Rc^2}.
\end{equation}

We match the interior solution with the Schwartzschild solution at the boundary of the star $r=R$. In addition we assume that the radial pressure vanishes at the boundary. Our boundary conditions are therefore
\begin{equation}\label{eq:boundary_cond} 
\alpha(r=R)= \ln \left(1-\frac{2GM}{Rc^2}\right), \quad 
\beta(r=R)= \ln \left(1-\frac{2GM}{Rc^2}\right)^{-1}, \quad
p_r(r=R) = 0.
\end{equation}
By use of Eqs. (\ref{eq:pot}), (\ref{eq:p_r_a}), the boundary conditions (\ref{eq:boundary_cond}) give the dimensionless parameters $a_0$, $a_1$ and $a_2$ with respect to the compactness of the star as follows
\begin{equation}
\label{eq:a_param} 
a_0(C) = \frac{1}{2}\frac{C}{1-C}, \quad  a_1 (C)=-(a_0 (C) +a_2 (C)),
\quad 
a_2(C)= \ln\left(1- C\right)^{-1}\,.
\end{equation}
Note also that since $C < 1$ we get
\begin{equation}
a_0>0,\quad a_1 <0,\quad a_2>0\,. 
\end{equation}

\begin{figure}[!tb]
	\begin{center}
		\subfigure[]{
			\label{fig:eos_r_C}
			\includegraphics[scale = 0.4]{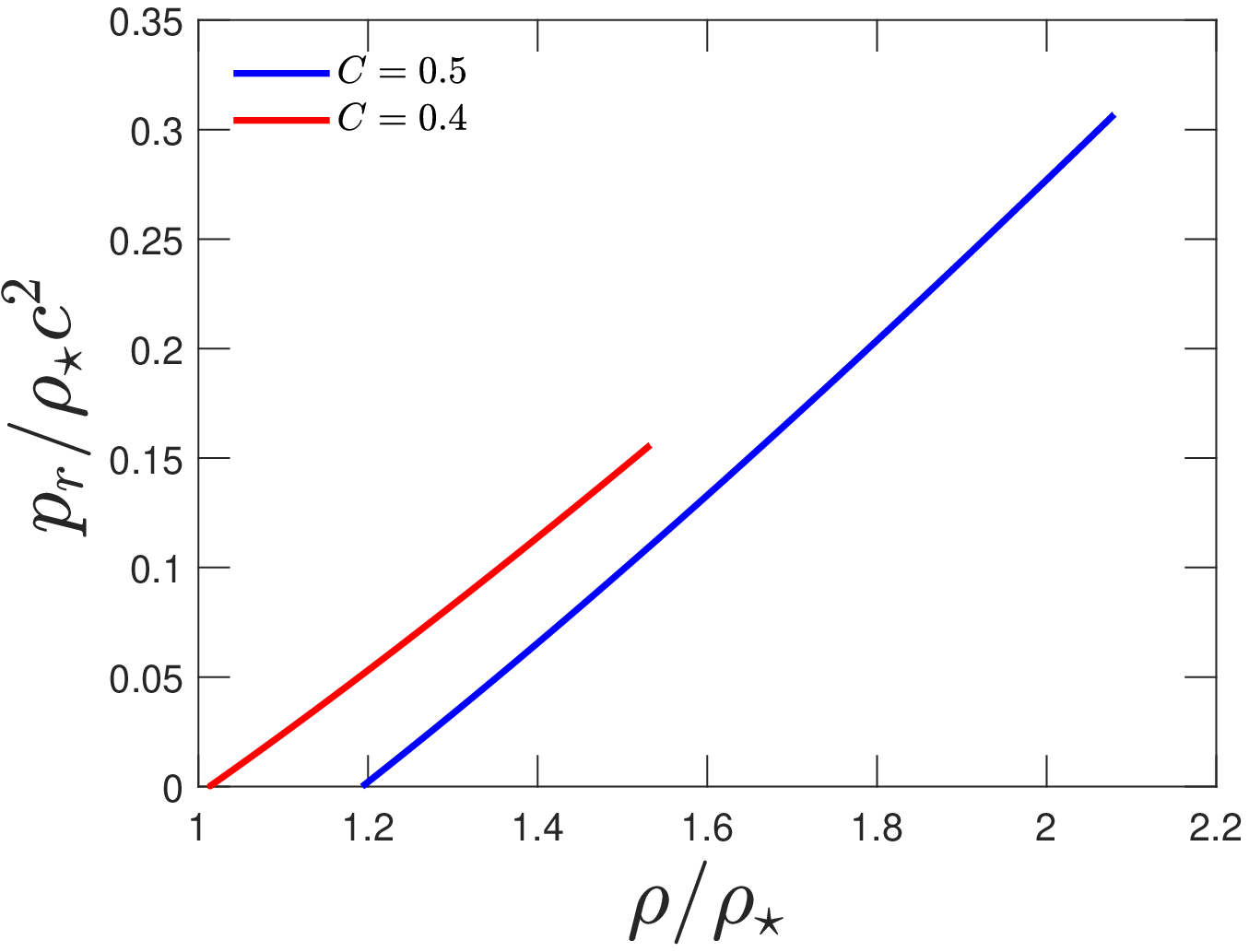}  }
		\subfigure[]{
			\label{fig:eos_t_C}
			\includegraphics[scale = 0.4]{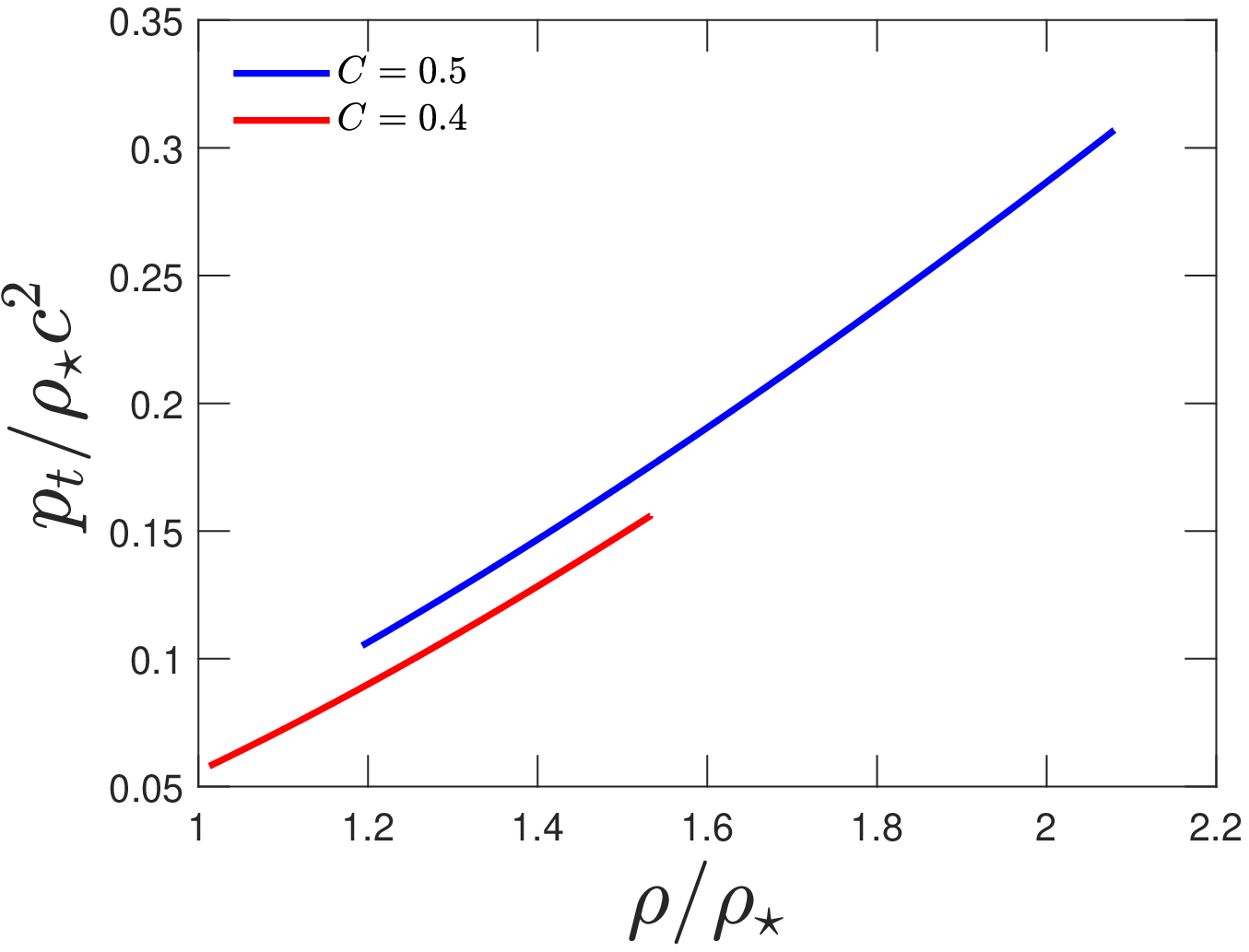}  }
		\\
		\subfigure[]{
	\label{fig:dpr_drho_C}
	\includegraphics[scale = 0.4]{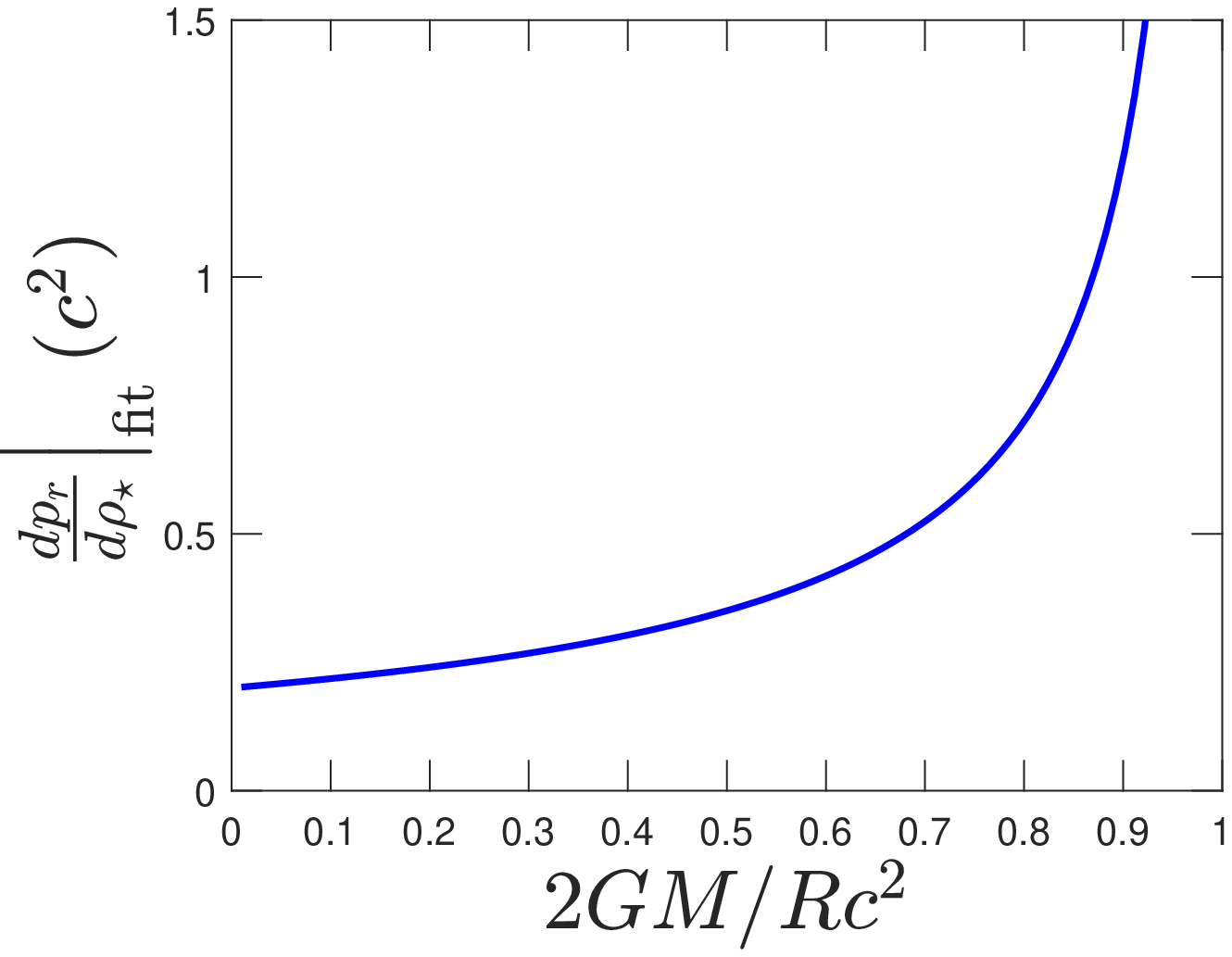}  }
\subfigure[]{
	\label{fig:dpt_drho_C}
	\includegraphics[scale = 0.4]{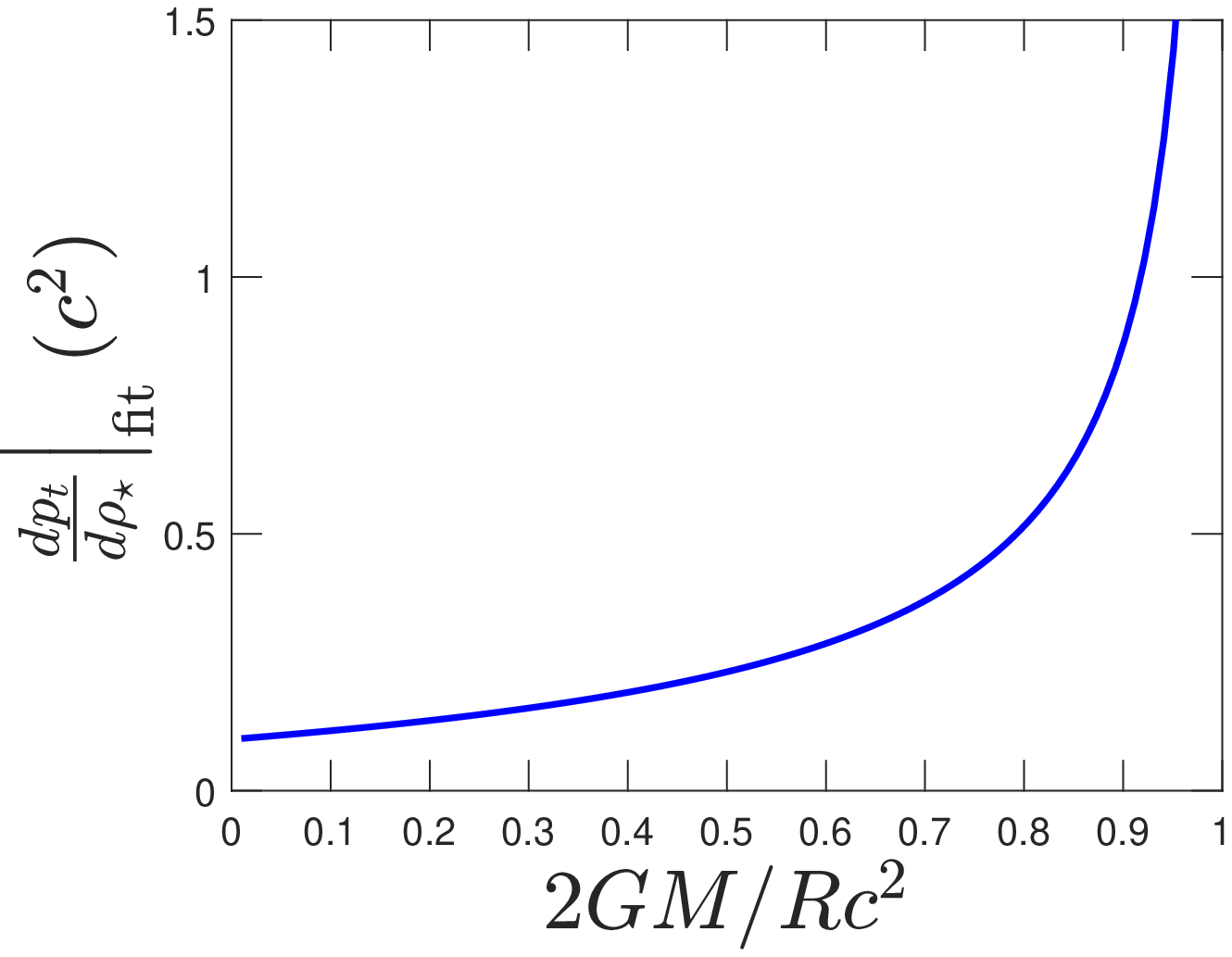}  }		
		\caption{In the upper panel is depicted the equation of state $p = p(\rho)$ for the radial and tangential pressures. In the lower panel we draw the slope of the linear fit of $p=p(\rho)$ for the radial and tangential pressures with respect to the compactness of the star. }
		\label{fig:eos_C}
	\end{center} 
\end{figure}

Thus, we parametrized any neutron star model in KB-spacetime with respect to the compactness of the star.
The quantities $\tilde{\rho} (x;C) $, $\tilde{p}_r(x;C) $, $\tilde{p}_t (x;C) $ are the same for stars with the same compactness $C$. In Figure \ref{fig:eos_C} is evident that the equation of state  $p_r=p_r(\rho)$, $p_t = p_t(\rho)$ is well fitted by a linear fit. In Figures \ref{fig:dpr_drho_C}, \ref{fig:dpt_drho_C} we plot the slope of the fit with respect to compactness. It is evident that causality is violated for sufficiently high compactness. We discuss in detail the constraints imposed by requiring that the solution is stable and physical in the next section.

\section{Constraints imposed by Physical and Stability Conditions}\label{sec:phys_an}

We impose the following reasonable physical requirements to a KB-spacetime (outlined in general very nicely in  \cite{2017EPJC...77..738I}). 

\begin{enumerate}[label=\roman*]
	\item The gravitational potentials $\exp (\alpha(r))$, $\exp (\beta (r))$ and the physical quantities $\rho(r)$, $p_r(r)$, $p_t(r)$ should be well defined in the center as well as regular and singularity free throughout the interior of the star. These properties are directly implied by Eqs. (\ref{eq:pot}), (\ref{eq:rho_a}), (\ref{eq:p_r_a}), (\ref{eq:p_t_a}).
	
	\item The anisotropy parameter should be positive throughout the whole interior of the star \cite{PhysRevD.70.024010,Mak:2003kw}. Indeed, we have that
	\begin{equation}\label{eq:Delta_pos}
	e^{a_2 x^2} - 1 - a_2 x^2 - a_0(a_2-a_0) x^4 = 
	\sum_{n=3}^\infty \frac{1}{n!}(a_2  x^2)^n
	+ \frac{1}{2}(a_2-a_0)^2  x^4 + \frac{1}{2} a_0^2  x^4 > 0.
	\end{equation}
	From Eq. (\ref{eq:Delta_a}) follows directly that $\Delta>0$ for all $a_0$, $a_2$, $r$. It should also be vanishing in the center. Indeed, we have
	\begin{equation}\label{eq:p_center}
	\tilde{p}_r(0)=
	\tilde{p}_t(0)=  2a_0-a_2.
	\end{equation}
	
	\item The energy density, the radial and tangential pressures should be positive at the center. The energy density at the center is
	\begin{equation}\label{eq:rho_0}
	\tilde{\rho}(0)=  3a_2 >0.
	\end{equation} 
	It is positive also throughout
	the stellar interior. Indeed
	\begin{equation}
	e^{a_2r^2}-1 + 2a_2r^2 = 
	1 + a_2r^2 + \sum_{n=2}^{\infty} \frac{1}{n!}(a_2 r^2)^n - 1 + 2a_2r^2  = 
	\sum_{n=2}^{\infty} \frac{1}{n!}(a_2 r^2)^n  + 3a_2r^2 >0
	\end{equation} 
	which, considering Eq. (\ref{eq:rho_a}), implies $\rho > 0$. 
	Regarding the radial and tangential pressure from Eq. (\ref{eq:a_param}) we get
	\begin{equation}\label{eq:a_0_2_cond}
	a_0 = a_2 f(C),\quad
	{\rm where }\,
	f(C) \equiv \frac{1}{2}\frac{C}{(1-C)\ln(1-C)^{-1}}\,.
	\end{equation}
	Since $C<1$ we have $f(C) > 1/2$. This is straightforward because $f(0) = 1/2$ and $df/dC > 0$. Thus
	\begin{equation}\label{eq:a02_ineq}
	a_0 > \frac{a_2}{2}.
	\end{equation}
	from which it follows $p_r(0)>0$, $p_t(0)>0$ where the central pressures are given in (\ref{eq:p_center}). 
	The radial pressure is positive also throughout the whole interior of the star since
	\begin{equation}\label{eq:ineq_1}
	1- e^{a_2 x^2} + 2a_0 x^2 = 
	\frac{(1-C)^{x^2} - (1-C)}{(1-C)^{x^2+1}} > 0,
	\end{equation}
	because $C<1$ and $x\leq 1$. Considering Eq. (\ref{eq:p_r_a}), it follows that $p_r>0$. Since the anisotropy parameter is positive, the tangential pressure is also positive throughout the interior of a star.

	\item The density and the pressures should be decreasing functions of $r$. We have 
	\begin{align}
	\label{eq:rho_prime}
	\tilde{\rho}{\,}^{\prime} &= 2
	\frac{ e^{-a_2 x^2}}{x^3} \left( -e^{a_2 x^2} + 1 + a_2 x^2  - 2a_2^2x^4 \right)
	\,, 
	\\
	\label{eq:p_r_prime}
	\tilde{p}_r{}^{\prime} &= 2
	\frac{ e^{-a_2 x^2}}{x^3} \left( e^{a_2 x^2} - 1 - a_2x^2 - 2a_0 a_2 x^4 \right)\,,
	\\
	\label{eq:p_t_prime}
	\tilde{p}_t{}^{\prime} &= 
	2 x e^{-a_2x^2}
	\left( a_0^2 + a_2^2 - 3a_0a_2 -  a_0 a_2 (a_0 - a_2) x^2 \right)\, ,
	\end{align}
	where prime denotes derivative with respect to $x$.
	We have 
	\begin{equation}\label{eq:ineq_2}
	-e^{a_2 x^2} + 1 + a_2 x^2  - 2a_2^2x^4 = 
	- \sum_{n=2}^\infty \frac{1}{n!} (a_2 x^2)^n
	- 2a_2^2 x^4 < 0\,,
	\end{equation}
	which, considering Eq. (\ref{eq:rho_prime}), implies $\rho ' < 0 $.
	Regarding now $p_r'$, we have by use of (\ref{eq:a_param})  that on the boundary the derivative is negative
	\begin{equation}
	e^{a_2 } - 1 - a_2 - 2a_0 a_2  = 
	\frac{ C - \ln (1-C)^{-1} }{1-C} < 0 
	\Rightarrow p_r'(x=1) < 0\, .
	\end{equation}
	Since we have proven that $p_r>0$, it is possible for $p_r$ to be increasing somewhere in the interval $x\in (0,1)$ if and only if there exists $x_0$ such that
	\begin{equation}\label{eq:p_r_dprime}
	\tilde{p}_r''(x_0) = 0 
	\Rightarrow 1 - e^{a_2 x_0^2} + 2a_0 x_0^2 = - 2a_0 x_0^2
	\,.
	\end{equation}
	However we have that 
	\begin{equation} 
	\tilde{p}_r > 0 \Rightarrow 1 - e^{a_2 x^2} + 2a_0 x^2 > 0\,,\quad \forall x\in (0,1)
	\end{equation}	
	and thus Eq. (\ref{eq:p_r_dprime}) is impossible. Since $p_r$ is positive in the interval $x\in [0,1)$, vanishes in the boundary $x=1$, and has no inflection point in the interval $(0,1)$ it follows that it is a decreasing function in $[0,1]$. The case of tangential pressure is more involved because there do exist inflection points of $p_t$. This occurs at high compactness and at low distance from the center, as depicted in Figure \ref{fig:p_t_prime_surface}. We calculate numerically that for $p_t$ being a monotonically decreasing function the compactness should satisfy that
	\begin{equation}\label{eq:C_constraint_pt}
	C < 0.94\, .
	\end{equation}

	\begin{figure}[!tb]
		\begin{center}
			\subfigure[]{
				\label{fig:p_t_prime_surface}
				\includegraphics[scale = 0.45]{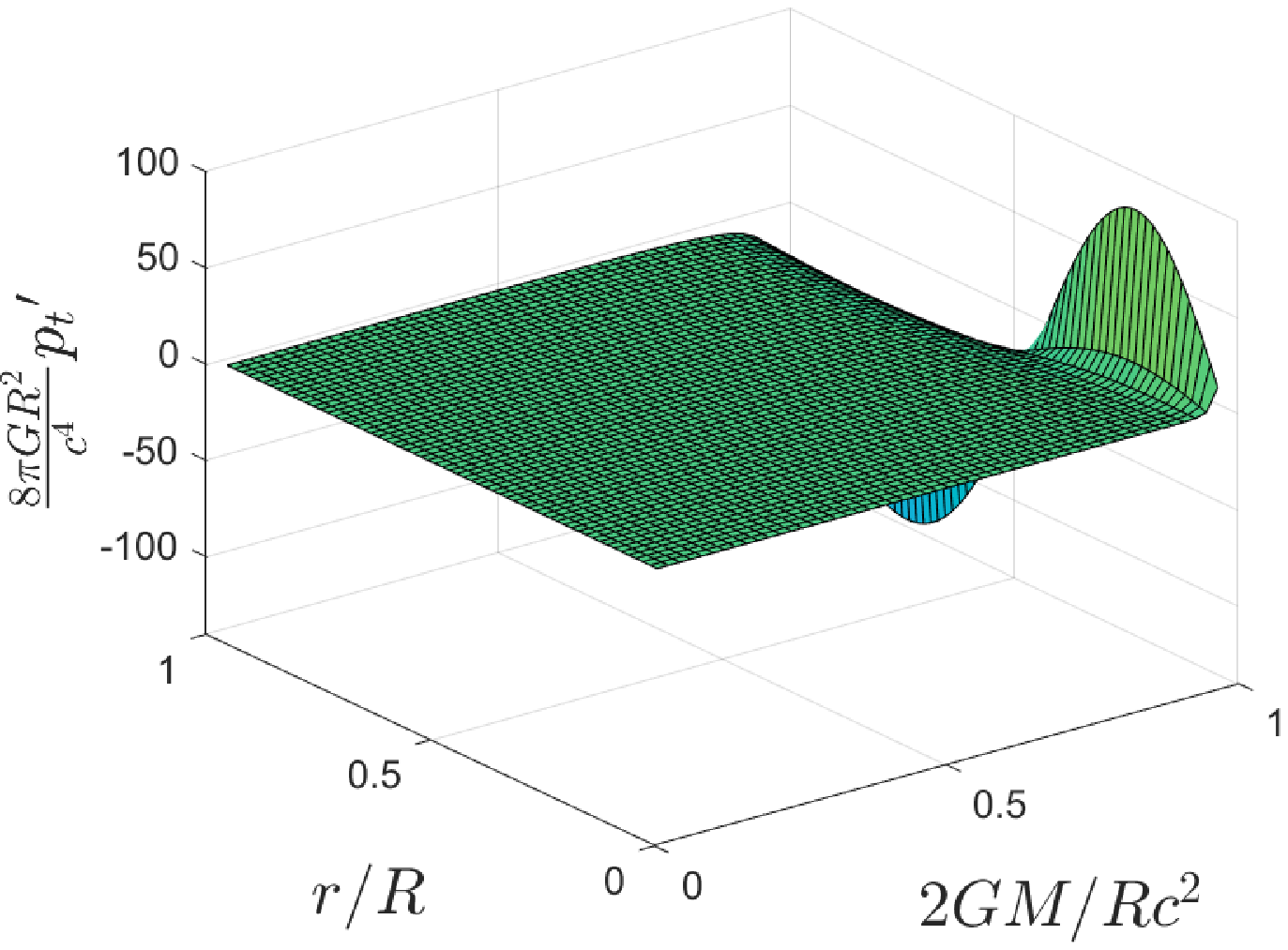}  }
			\subfigure[]{
				\label{fig:p_t_r}
				\includegraphics[scale = 0.45]{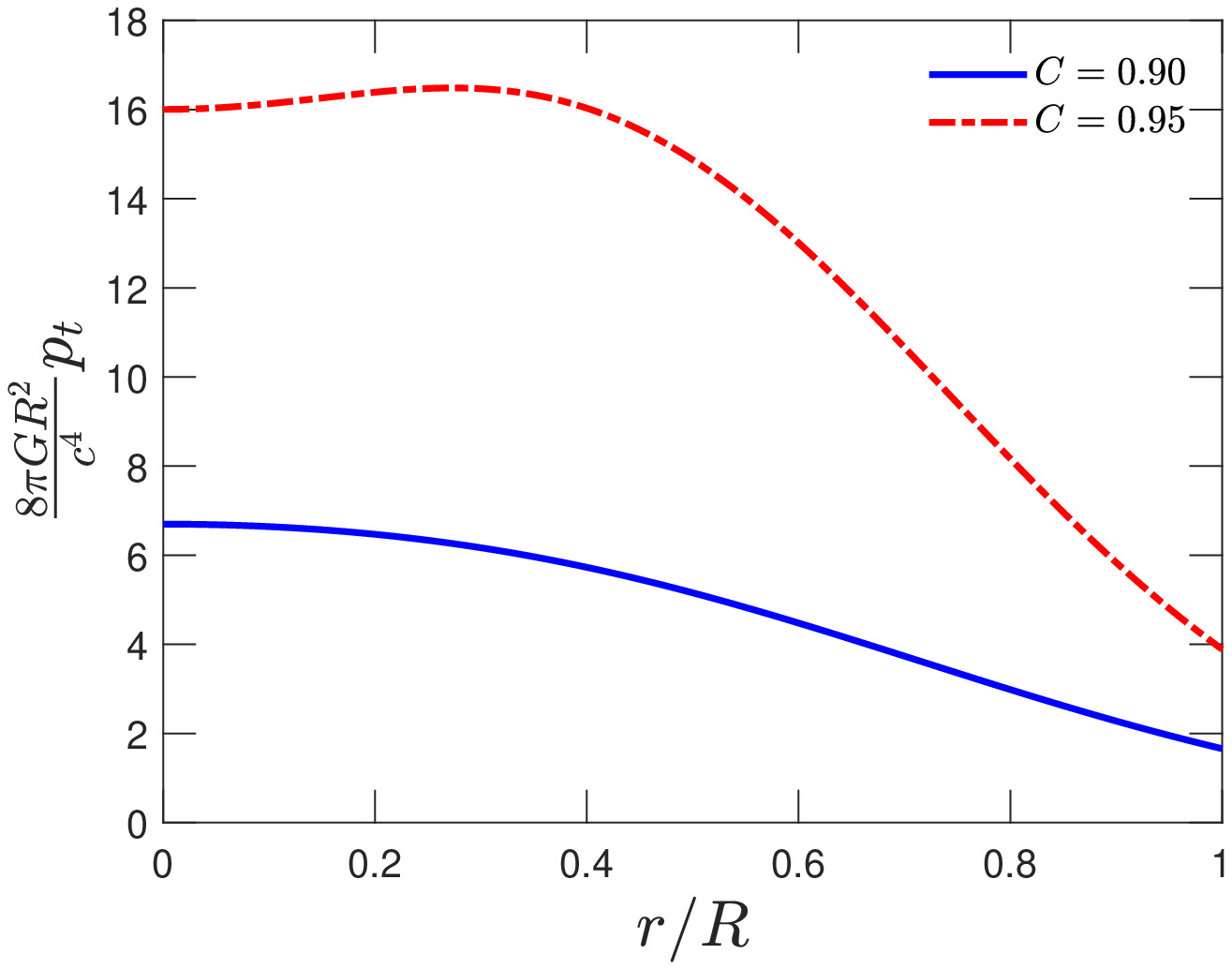}  }
			\caption{(a) The derivative of $p_t$ with respect to compactness $2GM/Rc^2$ of a star and distance $r/R$ from its center. It is evident that for high compactness values, $p_t$ is not monotonically decreasing. (b) The tangential pressure $p_t$ with respect to the distance from the center of the star for two compactness values.}
			\label{fig:p_t}
		\end{center} 
	\end{figure}

	\item We require that causality holds, namely that the speed of sound $v^2 = \frac{dp}{d\rho}$ is lower than the speed of light 
	\begin{equation}\label{eq:causality}
	0 \leq  v_r \leq c
	\,,\quad 
	0 \leq v_t \leq c
	,.
	\end{equation}
	The radial and transverse velocity of sound are
	\begin{align}
	\label{dso2}
	v_r{}^2 &=
	\frac{dp_r}{d\rho}=c^2 \frac{ 1 + a_2 x^2 + 2a_0 a_2 x^4 - e^{a_2 x^2}}{e^{a_2 x^2} - 1 -a_2 x^2+2a_2{}^2 x^4}, \\
	v_t{}^2
	&= \frac{dp_t}{d\rho}= c^2 \frac{x^4(3a_0a_2-a_0a_2{}^2 x^2+a_0{}^2a_2 x^2-a_2{}^2-a_0{}^2)}{e^{a_2 x^2} - 1 - a_2 x^2 + 2a_2{}^2 x^4}\, .
	\end{align}
	We find numerically that the inequalities (\ref{eq:causality}) impose constraints on maximum allowed compactness. The radial velocity imposes that $C < 0.86$ and the tangential velocity that $C<0.87$, therefore causality is satisfied if
	\begin{equation}\label{eq:C_constraint_caus}
	C < 0.86.
	\end{equation}
	These results are also in accordance with the linear fit of Figure \ref{fig:eos_C}.
	The fact that causality is not satisfied for sufficiently high compactness is depicted in Figure \ref{fig:causality}.
	
	\begin{figure}[!tb]
		\begin{center}
			\subfigure[Radial velocity of sound.]{
				\label{fig:v_r_surface}
				\includegraphics[scale = 0.45]{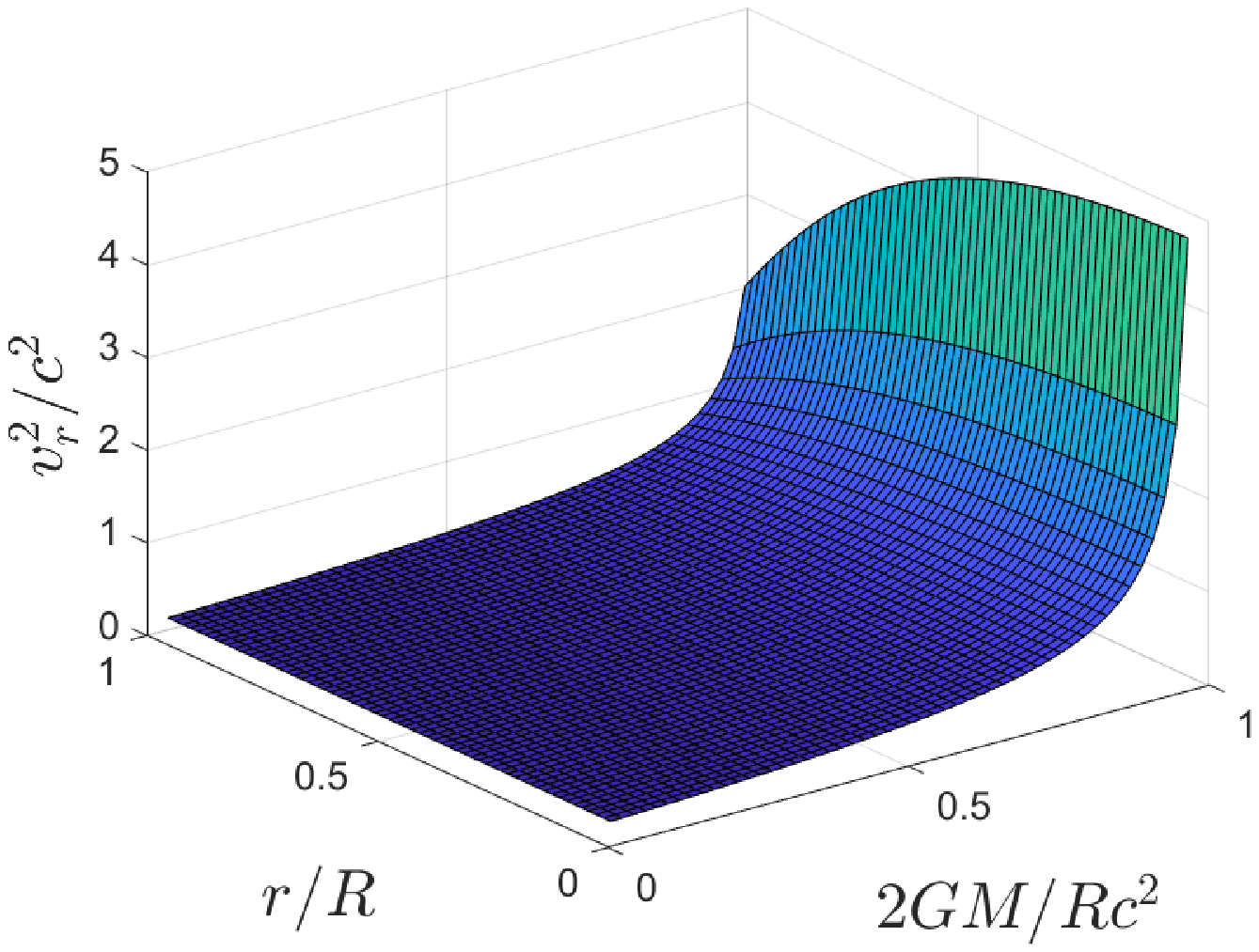}  }
			\subfigure[Tangential velocity of sound.]{
				\label{fig:v_t_surface}
				\includegraphics[scale = 0.45]{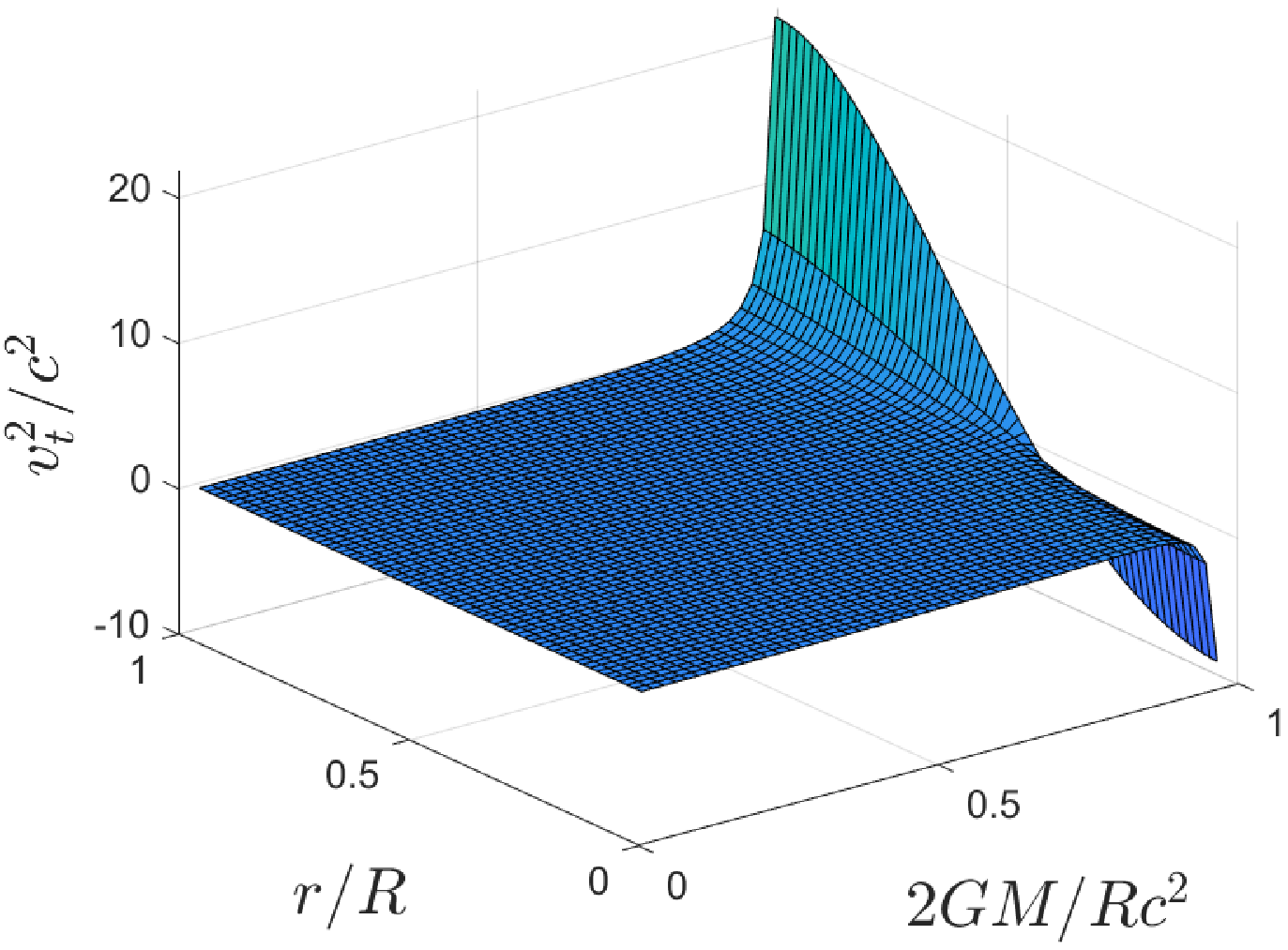}  }
			\caption{For sufficiently high compactness, causality is violated.}
			\label{fig:causality}
		\end{center} 
	\end{figure}

	\item We require stability against cracking which is satisfied if \cite{HERRERA1992206,2007CQGra..24.4631A}
	\begin{equation}\label{eq:cracking}
	0 < v_r{}^2-v_t{}^2 < c^2 \,. 
	\end{equation}
	This condition imposes an additional constraint on maximum allowed compactness
	\begin{equation}\label{eq:C_constraint_crack}
	C < 0.78.
	\end{equation}
	The fact that cracking stability is violated for sufficiently high compactness is depicted in Figure \ref{fig:cracking}.
	\begin{figure}[!tb]
		\begin{center}
			\subfigure[]{
				\label{fig:crack_surface}
				\includegraphics[scale = 0.45]{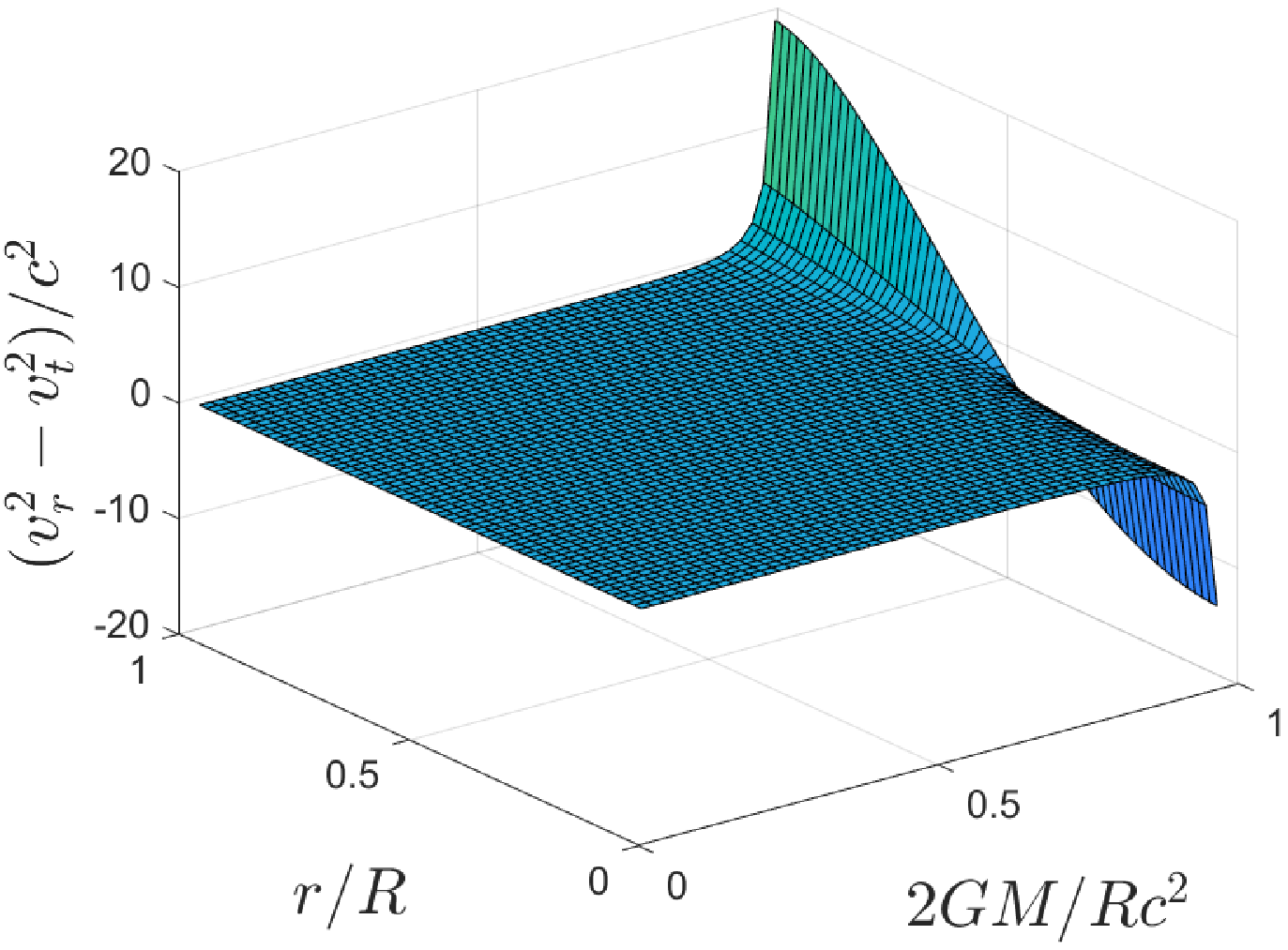}  }
			\subfigure[]{
				\label{fig:cracking_C}
				\includegraphics[scale = 0.45]{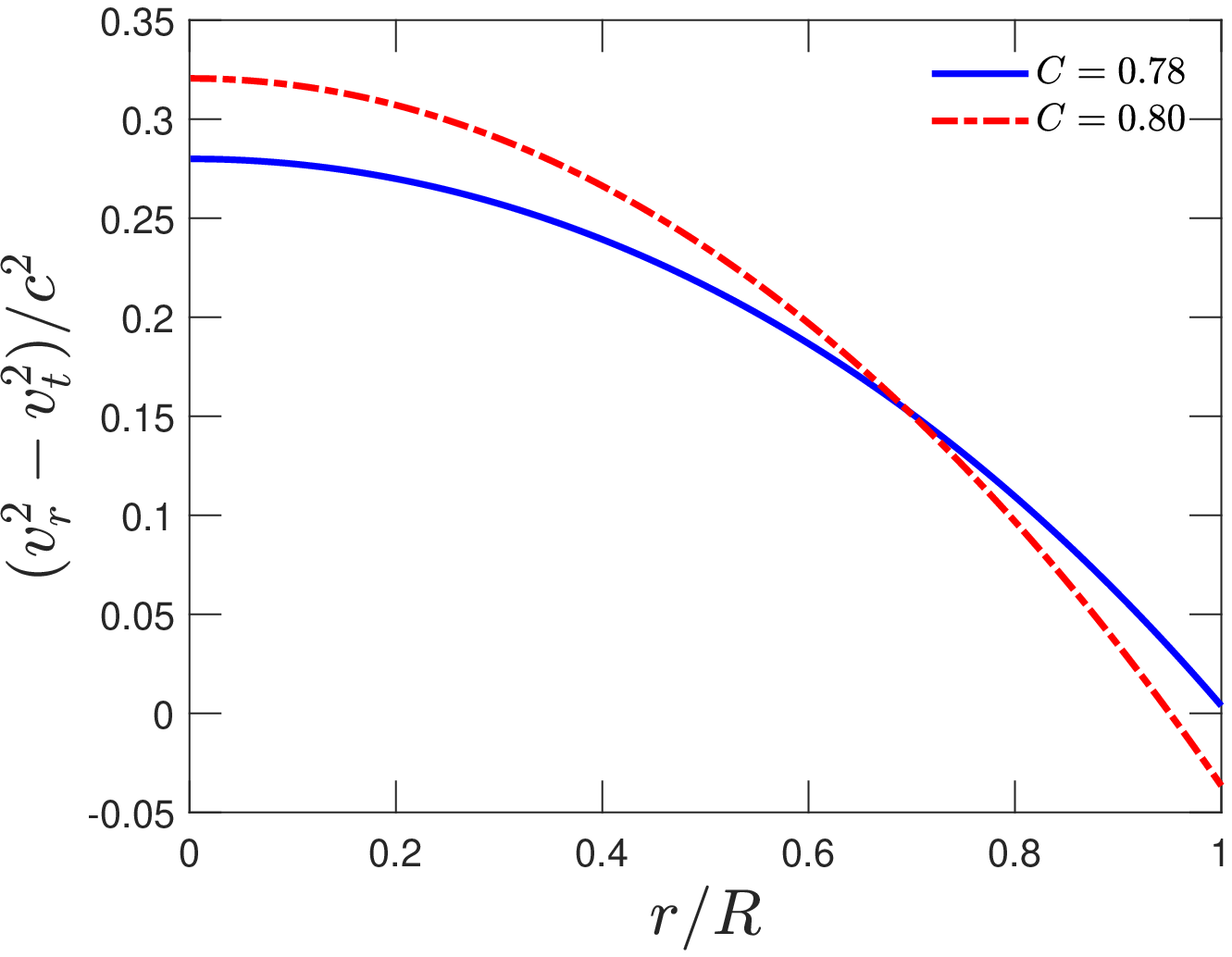}  }
			\caption{For sufficiently high compactness, stability against cracking as in condition (\ref{eq:cracking}) is violated.}
			\label{fig:cracking}
		\end{center} 
	\end{figure}

	\item We require that the strong energy condition (SEC) \cite{1988CQGra...5.1329K,2017EPJC...77..738I}  holds
	\begin{equation}\label{eq:SEC}
	\rho c^2 - p_r - 2p_t > 0\,.
	\end{equation} 
	It places an additional constraint to the maximum allowed compactness, which we calculate numerically to be
	\begin{equation}\label{eq:C_constraint_SEC}
	C \leq 0.715\,.
	\end{equation}
	In Figure \ref{fig:SEC} is depicted that indeed for sufficiently high compactness, SEC is violated.
	\begin{figure}[!tb]
		\begin{center}
			\subfigure[]{
				\label{fig:SEC_surface}
				\includegraphics[scale = 0.45]{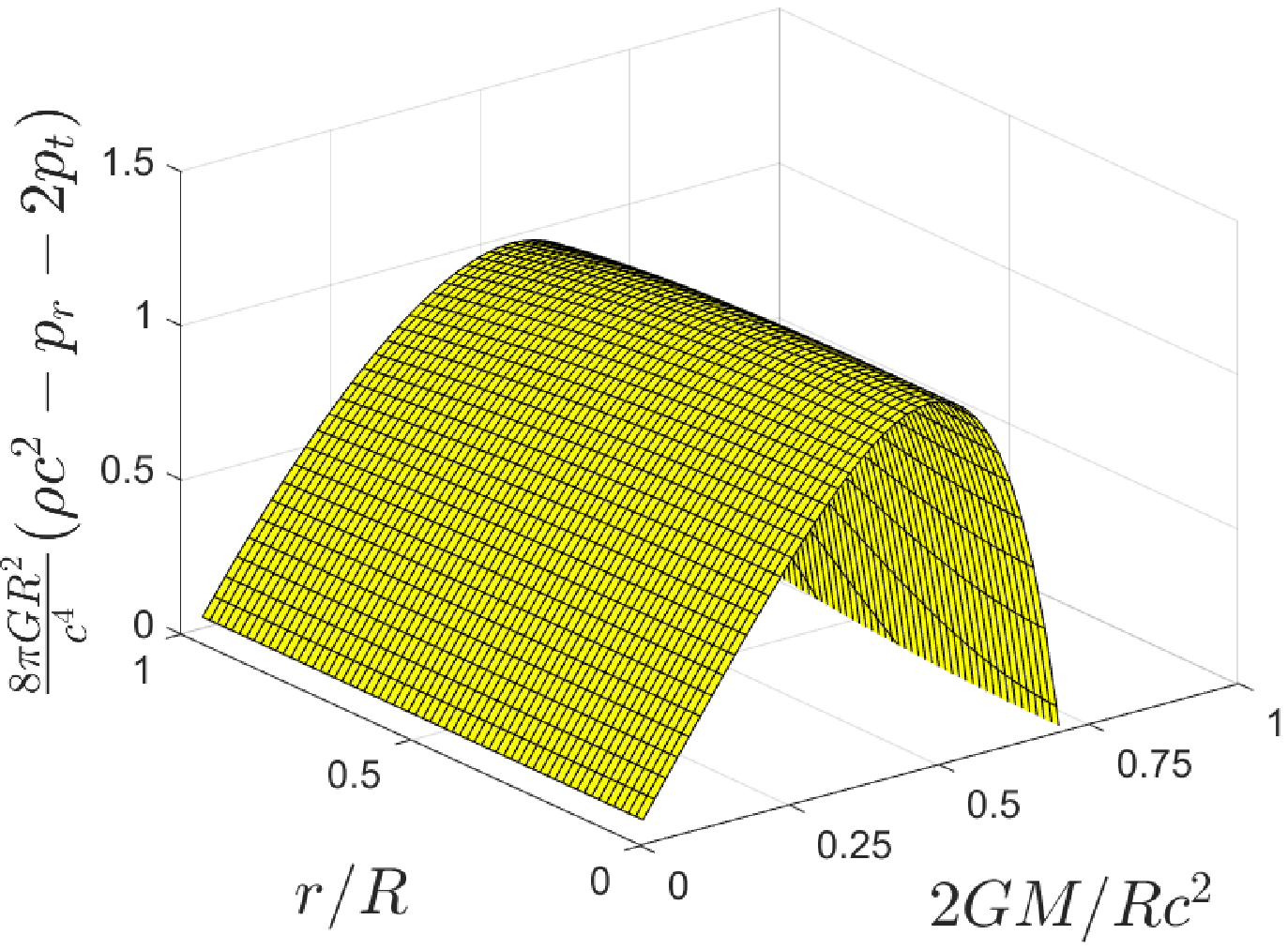}  }
			\subfigure[]{
				\label{fig:SEC_C}
				\includegraphics[scale = 0.45]{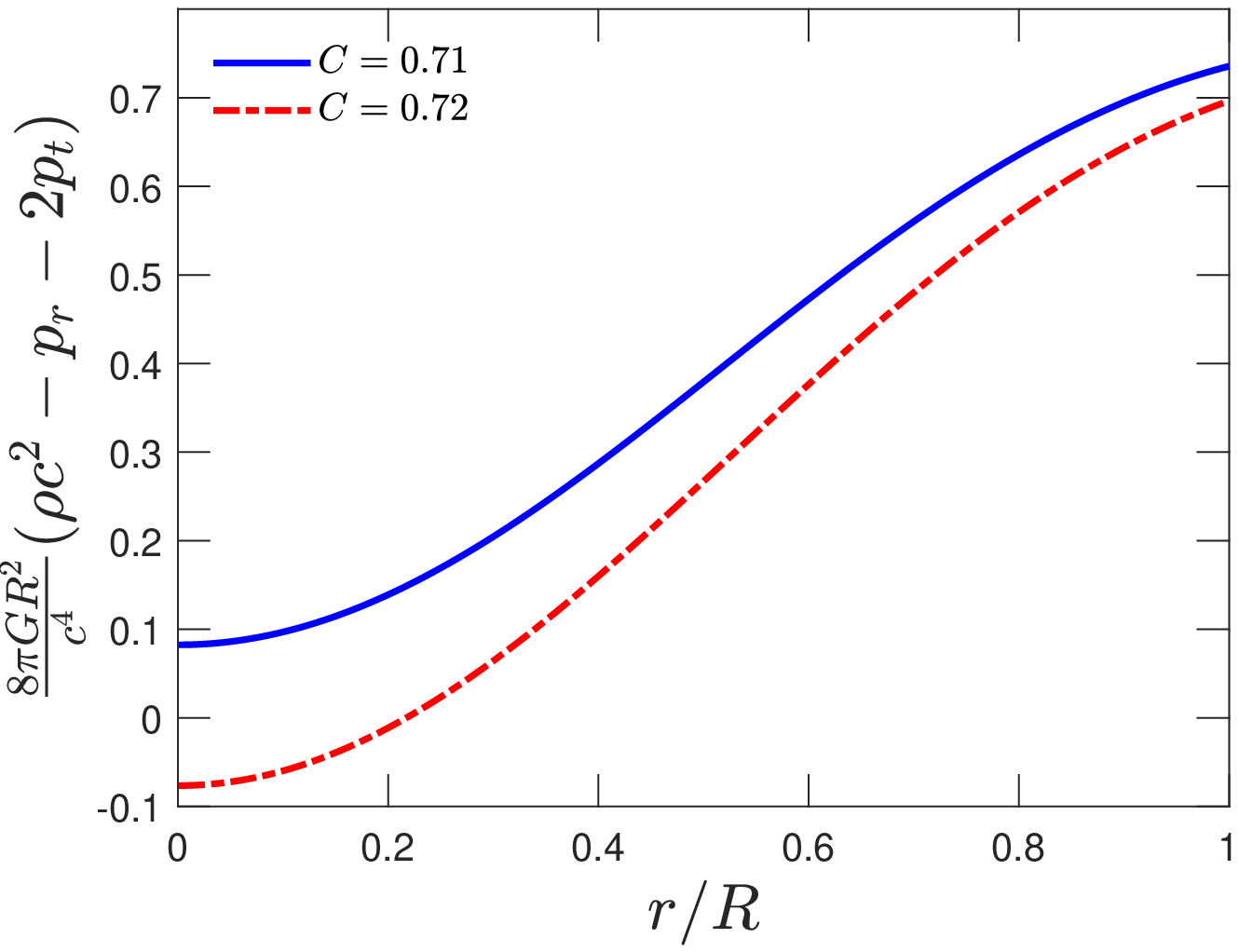}  }
			\caption{For sufficiently high compactness, SEC as in condition (\ref{eq:SEC}) is violated.}
			\label{fig:SEC}
		\end{center} 
	\end{figure}
	
	\item Our model satisfies the stability condition for the adiabatic index \cite{1994MNRAS.267..637C,1997PhR...286...53H}
	\begin{equation}\label{eq:gamma_cond}
	\Gamma \equiv \frac{\rho+ p_r}{p_r}\frac{dp_r}{d \rho} > \frac{4}{3}.
	\end{equation}
The adiabatic index may be written as
	\begin{equation}
	\Gamma (x;C) = \frac{2 x^2(a_0 + a_2)(1 + a_2  x^2 + 2 a_0 a_2  x^4 - e^{a_2 x^2} )}{(1 - e^{a_2 x^2} + 2a_0 x^2  )(e^{a_2 x^2} - 1 - a_2 x^2 + 2a_2{}^2 x^4)}.
	\end{equation}
	From inequalities (\ref{eq:ineq_1}), (\ref{eq:ineq_2}) it follows that $\Gamma > 0$. We verify numerically that $\Gamma$ is a monotonically increasing function for all $C\in[0,1]$ and $x\in [0,1]$, with $a_0=a_0(C)$, $a_2=a_2(C)$ as in Eq. (\ref{eq:a_param}). This is depicted in Figure (\ref{fig:Gamma_surface}). 
	We also have that
	\begin{equation}\label{eq:Gamma_0}
	\Gamma (x=0) = \frac{4}{5}\frac{(\frac{a_0}{a_2}+1)(2\frac{a_0}{a_2}+\frac{1}{2})}{2\frac{a_0}{a_2}-1} > \frac{7}{5}\left(1+ \frac{2\sqrt{6}}{7}\right),
	\end{equation}
	where $\frac{a_0}{a_2} > \frac{1}{2} $ as in (\ref{eq:a02_ineq}).
	The inequality (\ref{eq:Gamma_0}) follows from the monotonicity of the function
	\begin{equation}
	h(z) = \frac{4}{5}\frac{(z+1)(2z+\frac{1}{2})}{2z-1}	
	\end{equation}
	in the interval $z>\frac{1}{2}$. The function $h(z)$ presents a minimum at $z_{\rm min} = \frac{1}{2}+\frac{\sqrt{6}}{4}$. It is 
		$h(z_{\rm min}) =  \frac{7}{5}\left(1+ \frac{2\sqrt{6}}{7}\right)$.
It follows that $\Gamma > 4/3$ for all $x$ and $C$ in the interval $[0,1]$.
	
	\begin{figure}[!tb]
		\begin{center}
			\includegraphics[scale = 0.5]{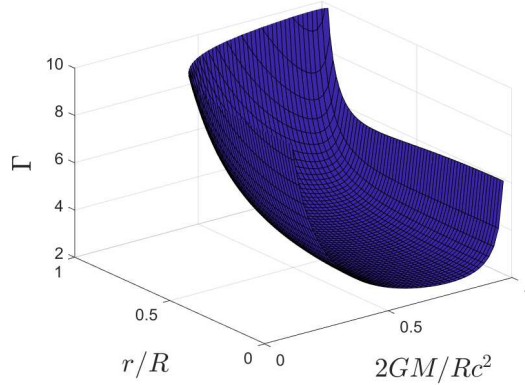}  
			\caption{The adiabatic index $\Gamma$ with respect to compactness and distance from the center.}
			\label{fig:Gamma_surface}
		\end{center} 
	\end{figure}
	
\end{enumerate}

To conclude, we find that conditions (i), (ii), (iii), (viii) are all satisfied by KB-spacetime for any compactness value. Conditions (iv), (v), (vi), (vii) impose constraints to compactness given by (\ref{eq:C_constraint_pt}), (\ref{eq:C_constraint_caus}), (\ref{eq:C_constraint_crack}), (\ref{eq:C_constraint_SEC}), respectively. Combining these constraints we get that it should hold
\begin{equation}\label{eq:C_constraint}
C \leq 0.715,
\end{equation}
for a stable, physical solution within a KB-spacetime.

\section{Observational Data}\label{sec:obs}

\begin{small}
	\begin{table}[tbp]
		\begin{center}
			\begin{tabular}{l | c c c | c c}
				\toprule
				Pulsar & 
				$M (M_\odot)$ &
				$R (\rm km)$ &
				Reference &
				$\rho_R (10^{14}{\rm gr/cm^3})$ &
				$\frac{d p_r}{d\rho}(c^2)$ 
								\\
				\midrule
				\multicolumn{6}{c}{Millisecond Pulsars with White Dwarf Companions}
				\\
				\midrule
				J0437-4715 &
				$1.44^{+0.07}_{-0.07}$ &
				$13.6^{+0.9}_{-0.8}$ &
				\cite{2016MNRAS.455.1751R,2019MNRAS.490.5848G} &
				$2.4$ &
				$0.27$
				\\ [1.5ex]
				J0030+0451 &
				$1.44^{+0.15}_{-0.16}$ &
				$13.02^{+1.24}_{-1.06}$ &
				\cite{2019ApJ...887L..24M} &
				$2.7$ &
				$0.28$
								\\ [1.5ex]
				&
				$1.34^{+0.15}_{-0.16}$ &
				$12.71^{+1.14}_{-1.19}$ &
				\cite{2019ApJ...887L..21R} &
				$2.7$ &
				$0.27$		
\\
\midrule
				\multicolumn{6}{c}{Gravitational-wave Signals}
				\\
				\midrule
				LIGO constraints
&
$1.4$ &
$12.9^{+0.8}_{-0.7}$ &
\cite{2020ApJ...896L..44A} &
$2.7$ &
$0.27$	
\\ [1.5ex]
				GW170817-1 &
				$1.45^{+0.09}_{-0.09}$ &
				$11.9^{+1.4}_{-1.4}$ &
				\cite{PhysRevLett.121.161101} &
				$3.5$ &
				$0.29$
				\\ [1.5ex]
				GW170817-2 &
				$1.27^{+0.09}_{-0.09}$ &
				$11.9^{+1.4}_{-1.4}$ &
				\cite{PhysRevLett.121.161101} &
				$3.1$ &
				$0.27$	
				\\
				\midrule
				\multicolumn{6}{c}{Quiescent Low-mass X-ray Binaries}
\\
\midrule
X7 &
$1.4$ &
$14.5^{+1.8}_{-1.6}$ &
\cite{2006ApJ...644.1090H} &
$2.0$ &
$0.26$
\\ [1.5ex]
M13 &
$1.38^{+0.08}_{-0.23}$ &
$9.95^{+0.24}_{-0.27}$ &
\cite{2007ApJ...671..727W} &
$5.6$ &
$0.31$	
\\
\midrule
				\multicolumn{6}{c}{Pulsars Presenting Thermonuclear Bursts}
				\\
				\midrule
				4U 1724-207 &
$1.81^{+0.25}_{-0.37}$ &
$12.2^{+1.4}_{-1.4}$ &
\cite{2016ApJ...820...28O} &
$3.9$ &
$0.32$
\\ [1.5ex]
4U 1820-30 &
$1.46^{+0.21}_{-0.21}$ &
$11.1^{+1.8}_{-1.8}$ &
\cite{2016ApJ...820...28O} &
$4.3$ &
$0.30$
\\ [1.5ex]
SAX J1748.9-2021 &
$1.81^{+0.25}_{-0.37}$ &
$11.7^{+1.7}_{-1.7}$ &
\cite{2016ApJ...820...28O} &
$4.4$ &
$0.33$
\\ [1.5ex]
				EXO 1745-268 &
$1.65^{+0.21}_{-0.31}$ &
$10.5^{+1.6}_{-1.6}$ &
\cite{2016ApJ...820...28O} &
$5.5$ &
$0.33$
\\ [1.5ex]
				4U 1608-52 &
$1.57^{+0.30}_{-0.29}$ &
$9.8^{+1.8}_{-1.8}$ &
\cite{2016ApJ...820...28O} &
$6.4$ &
$0.34$
\\  [1.5ex]
				KS 1731-260 &
$1.61^{+0.35}_{-0.37}$ &
$10.0^{+2.2}_{-2.2}$ &
\cite{2016ApJ...820...28O} &
$6.2$ &
$0.34$
\\ 
				\bottomrule
			\end{tabular}
		\end{center} 
		\caption{The boundary density $\rho_R$ and slope of the $p_r(\rho)$ linear fit within our model for several observational data of mass $M$ and radius $R$ of pulsars. The ``LIGO constraints'' label refers to the radius constraints on a canonical mass neutron star obtained from GW170817, GW190814 in \cite{2020ApJ...896L..44A}.}
		\label{tab:density}
	\end{table}
\end{small}

Independent mass measurements of neutron stars are relatively easy. Their fast pulsation is ideal for timing measurements and since the pulsar is often  a member of a binary system, this precise timing can be used to measure its orbital motion with astonishing precision \cite{2016ARA&A..54..401O}. On the other hand the independent radius measurement, depending on the thermal emmision of the stellar surface, is much more difficult.

Rotation-powered millisecond pulsars allow for the measurement of radius based on thermal emmision in soft X-rays \cite{2016ARA&A..54..401O,2019MNRAS.490.5848G}. The mass of the pulsar can be measured independently by radio-timing measurements. There exist two such pulsars, PSR J0437-4715 (we shall call J0437) and PSR J0030+0451 (we shall call J0030), for which  there exist very recent independent and reliable measurements from  NICER  \cite{2019ApJ...887L..25B,2019ApJ...887L..24M}.

\begin{figure}[!tb]
	\begin{center}
		\includegraphics[scale = 0.6]{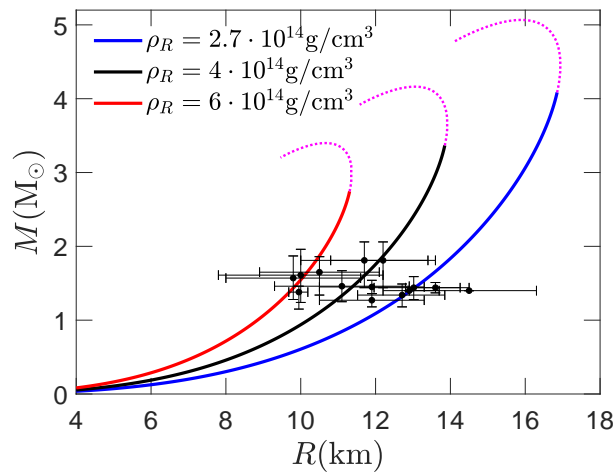}  
		\caption{The mass-radius curve for three different boundary conditions $\rho_R$ along with $(R,M)$ points of all pulsars of Table \ref{tab:density}. }
		\label{fig:M-R}
	\end{center} 
\end{figure}

In addition to J0437 and J0030 we consider the gravitational-wave signals GW170817\cite{TheLIGOScientific:2017qsa,PhysRevLett.121.161101} and GW190814 \cite{2020ApJ...896L..44A}. GW170817 was the first detection of coalescence of two neutron stars \cite{TheLIGOScientific:2017qsa}. The subsequent analysis of LIGO/Virgo collaboration \cite{PhysRevLett.121.161101} constrained significantly the radii and masses of the two neutron stars. GW190814 was the first detection of an object within the mass-gap \cite{2020ApJ...896L..44A} with mass $M = 2.6{\rm M}_\odot$. If this component is a neutron star and not a black hole the equation of state is further constrained. 

We supplement our analysis with data regarding two quiescent low-mass X-ray binaries \cite{2006ApJ...644.1090H,2007ApJ...671..727W} and six pulsars, members of low-mass X-ray binaries, that present thermonuclear X-ray bursts. For such pulsars it is possible to get correlated $M-R$ constraints \cite{2006Natur.441.1115O,2016ARA&A..54..401O} independent from assumptions regarding the equation of state. 

In Table \ref{tab:density}	we see that the boundary density of all pulsars within a KB-spacetime lies in the range $(2.5-6.5)\cdot 10^{14}{\rm gr/cm^3}$. These values are consistent with a neutron core. 
All pulsars are consistent with a linear equation of state as already shown in Figure \ref{fig:eos_C}. The corresponding slope is depicted in the last column of Table \ref{tab:density}.

	\begin{figure}[!tb]
	\begin{center}
		\subfigure[]{
			\label{fig:M-R_max}
			\includegraphics[scale = 0.45]{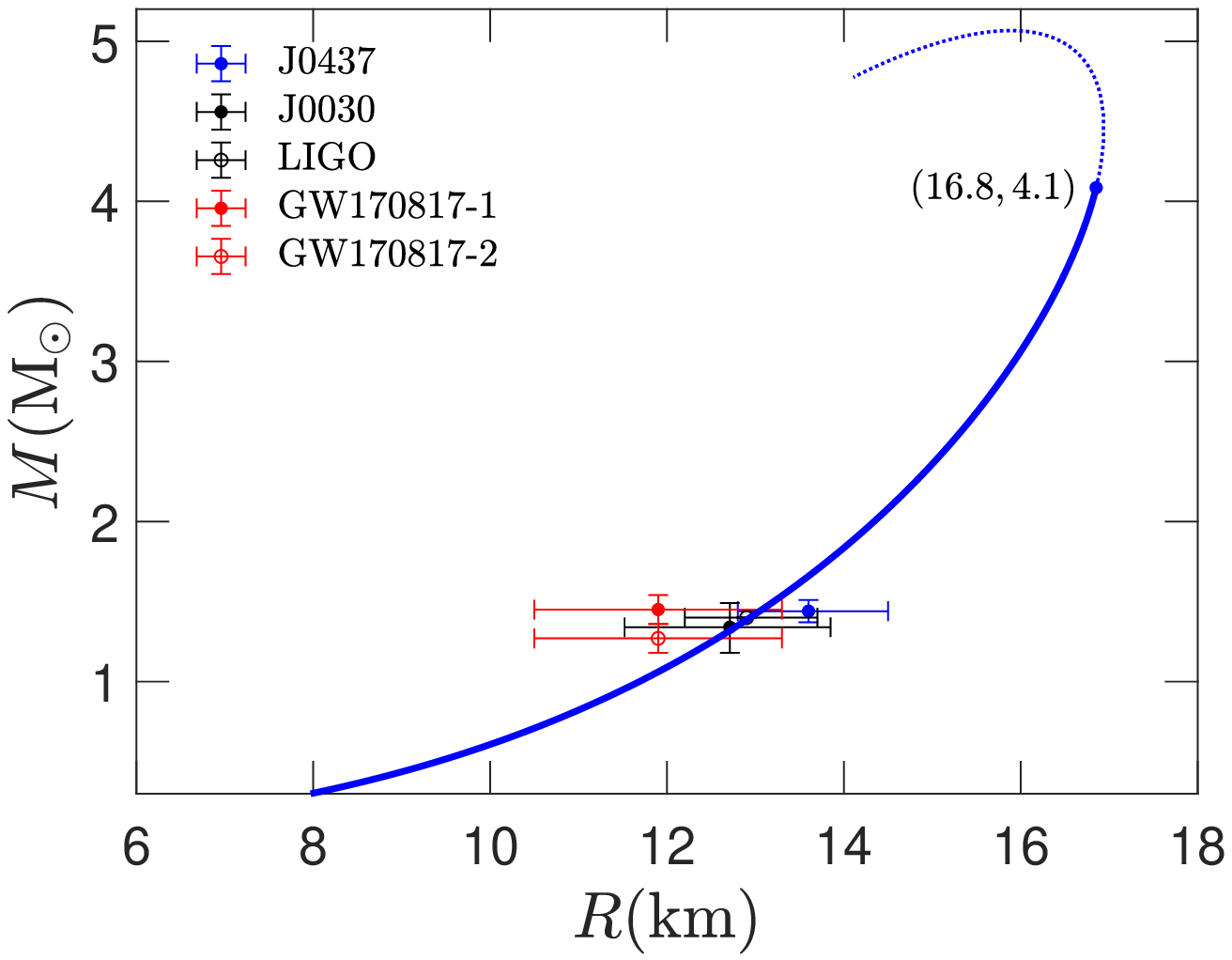}  }
		\subfigure[]{
			\label{fig:C-R_max}
			\includegraphics[scale = 0.45]{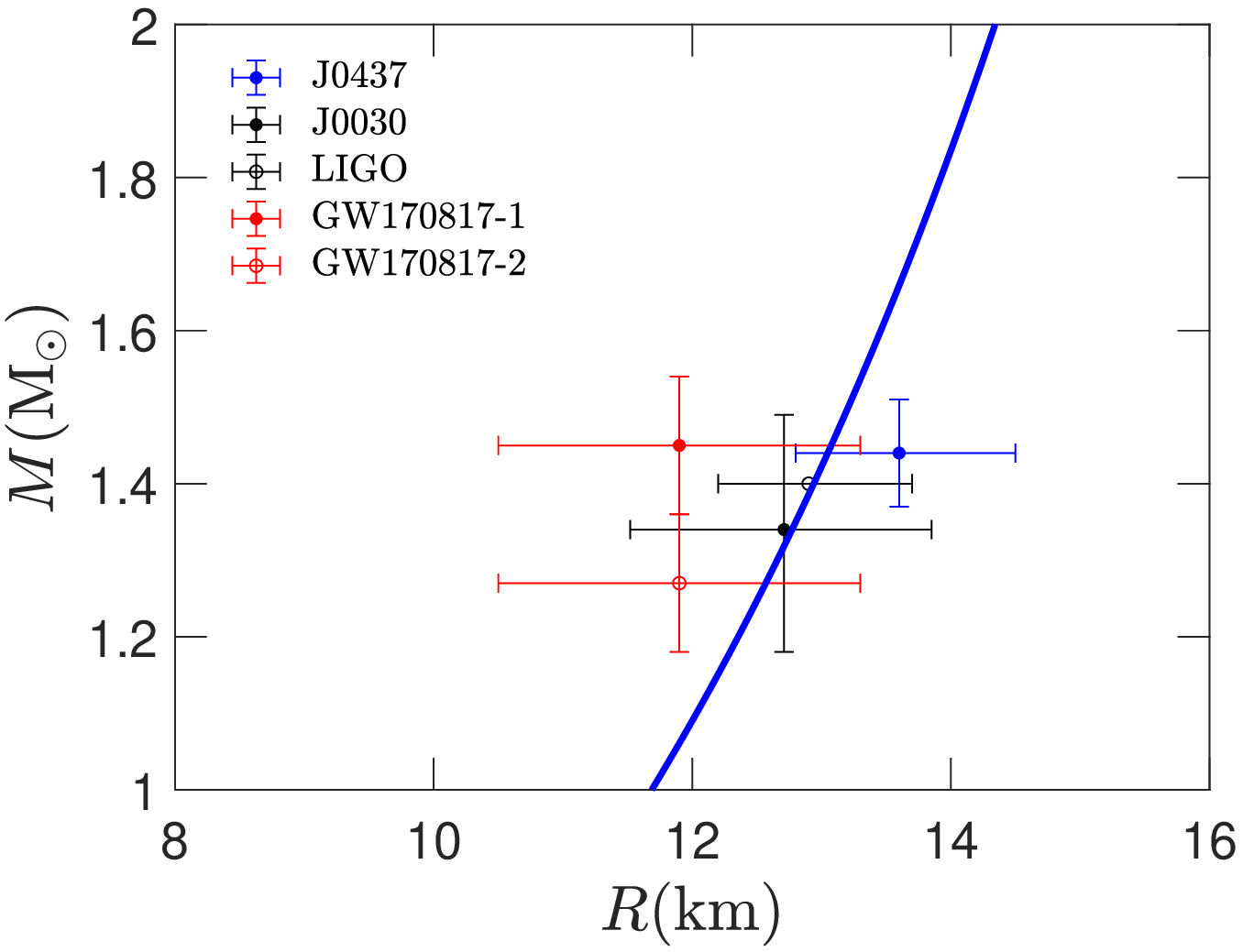}  }
		\caption{The curve corresponds to equilibrium solutions with the same boundary condition $\rho_R = 2.7\cdot 10^{14}{\rm gr/cm^3}$. The solid branch corresponds to stable, physical solutions and the dotted branch represents unstable or nonphysical solutions. The maximum mass is $4.1M_\odot$ at radius $16.8{\rm km}$ with corresponding compactness $0.71$. On the right panel we focus on the area including the most recent data from NICER and LIGO. The ``LIGO'' label refers to the radius constraints on a canonical mass neutron star obtained from GW170817, GW190814 in \cite{2020ApJ...896L..44A}. }
		\label{fig:MC-R}
	\end{center} 
\end{figure}

In Figure \ref{fig:M-R} is plotted the mass-radius curve with a certain boundary condition. It is evident that the most recent and reliable independent data from NICER and LIGO are consistent with the same boundary density which amazingly turns out to equal the nuclear saturation density $\rho_{\rm sat} = 2.7\cdot 10^{14}{\rm gr/cm^3}$. 
This value perfectly describes a neutron core since typically it designates core's boundary. 
Furthermore, anisotropies can grow down to this density. In Ref. \cite{1972NPhS..236...37C} is argued that solidification occurs at the nuclear density $\rho_{\rm solid}= 2.8\cdot 10^{14}{\rm gr/cm^3}$, and similar results are obtained in  \cite{1975ARA&A..13..335C} which predict $\rho_{\rm solid}= 3.7\cdot 10^{14}{\rm gr/cm^3}$. Even if solidification occurs at higher densities as predicted by other studies \cite{1973PhRvL..30..999C,1973NPhS..243..130S}, anisotropic superfluidity of nuclear matter appears at low densities such as $\rho = 1.5\cdot 10^{14} {\rm gr/cm^3}$ \cite{1970PhRvL..24..775H}. 

The mass-radius curve with the boundary condition $\rho_R = \rho_{\rm sat}$, which fits well NICER and LIGO data, is depicted in Figure \ref{fig:MC-R}). We have calculated in previous section the maximum allowed compactness for a physical anisotropic core within a KB-spacetime to be $C_{\rm max} = 0.71$. This limits the maximum mass to
$M_{\rm max} = 4.1M_\odot$
at 
$R = 16.8{\rm km}$. 

Our results does not exclude the possibility that the secondary component of GW190814 with mass $M=2.6{\rm M}_\odot$ is an anisotropic neutron star. The condition $C<0.71$ implies that it should have a radius $R>10.8{\rm km}$. The corresponding boundary density should be lower than $\rho_R< 6.6\cdot 10^{14}{\rm g}/{\rm cm}^3$.

\section{Conclusions}

We parametrized any neutron star model in KB-spacetime with respect to the compactness of the star. This description is given by Eqs. (\ref{eq:rho_a}), (\ref{eq:p_r_a}), (\ref{eq:p_t_a}), (\ref{eq:a_param}).
Requiring that any KB-spacetime model of neutron stars is physical and the solution stable constraints the compactness to a maximum value
\begin{equation}\label{eq:C_bound}
\frac{2GM}{Rc^2} < 0.71.
\end{equation}
This is significantly more strict than the bound, $0.95$, obtained for general anisotropic stars in Ref. \cite{Ivanov_2002} and is in contrast to suggestions that anisotropic compact stars can be arbitrarily compact \cite{GLEISER_2004,Bohmer2006}.

 The equations of state in a general KB-spacetime, depicted in Figure \ref{fig:eos_C}, are well fitted by a linear fit.  
A KB-spacetime fits observational data obtained from numerous pulsars of Table \ref{tab:density} with boundary density in the range $\rho_R \sim (2.5-6.5)\cdot 10^{14}{\rm gr/cm^3}$ consistent with an anisotropic neutron core. In Figure \ref{fig:M-R} we have calculated the mass-radius curves. An additional direct indication that KB-spacetime is realistic for anisotropic neutron stars is that the most recent data from NICER and LIGO are well fitted with boundary density which equals precisely the nuclear saturation density!

The mass-radius curve with boundary density that equals to the nuclear saturation density presents a mass maximum 
\begin{equation}
M < 4.1M_\odot
\end{equation}
when the bound (\ref{eq:C_bound}) to compactness is also taken into account. This limits the maximum allowed mass for any consistent compact star model in KB-spacetime. This result is also in accordance with the estimated maximum mass in general spacetimes by Heintzmann \& Hillebrandt \cite{1975A&A....38...51H}, who use semi-realistic equations of state.

Our analysis predicts further that anisotropic neutron stars may populate partially the mass gap $(2.5-5)M_\odot$ regarding compact objects. An observation of a neutron star with $M>2.5M_\odot$ will be an indication it is composed of an anisotropic core. Especially, the possibility that the secondary component of the gravitational-wave signal GW190814 is a neutron star with an anisotropic core requires further investigation.

\bibliography{2020_Roupas_Nashed_ANS_KB-ST}
\bibliographystyle{myunsrt}

\end{document}